\documentclass[article,groupedaddress,reprint,floats,nofootinbib,preprintnumbers]{revtex4-2}
\usepackage[final]{graphicx} 
\usepackage{float}
\usepackage{amsmath}
\usepackage{epstopdf}
\usepackage{booktabs}
\usepackage[normalem]{ulem} 
\usepackage[compat=1.1.0]{tikz-feynman}
\usepackage[breaklinks=true]{hyperref}
\usepackage{lipsum}
\hypersetup{
  colorlinks=true,
  linkcolor=magenta,
  citecolor=blue,
  urlcolor=magenta
}
\usepackage{array,tabularx,mathrsfs}
\usepackage{ifthen}
\usepackage{amsfonts}
\usepackage{amssymb}
\usepackage{slashed}

\newcommand{\magenta}[1]{\color{magenta} #1 \color{black}}

\newcommand{\be}{\begin{equation}}
\newcommand{\ee}{\end{equation}}
\newcommand{\LL}{\mathscr{L}}

\newcommand{\hc}{\text{h.c.}}
\newcommand{\ov}[1]{\overline{#1}}
\newcommand{\nn}{\nonumber}
\newcommand{\vev}[1]{\langle #1 \rangle}

\newcommand{\TeV}{\text{TeV}}
\newcommand\insim{\mathrel{%
  \ooalign{\raise0.2ex\hbox{$\in$}\cr\hidewidth\raise-0.8ex\hbox{\scalebox{0.9}{$\sim$}}\hidewidth\cr}}}

\begin{document}
\preprint{\magenta{IFT-UAM-CSIC-23-104}}
%

\title{\boldmath
Early Universe hypercharge breaking and neutrino mass generation
}

\author{S.~L\'opez-Zurdo}
\email[]{sergio.lopezz@estudiante.uam.es}
\author{A.~Lozano-Onrubia}
\email[]{alvaro.lozano.onrubia@csic.es}
\author{L.~Merlo}
\email[]{luca.merlo@uam.es}
\author{J.~M.~No}
\email[]{josemiguel.no@uam.es}
\affiliation{Departamento de F\'isica Te\'orica and Instituto de F\'isica Te\'orica UAM/CSIC,
Universidad Aut\'onoma de Madrid, Cantoblanco, 28049, Madrid, Spain}


\begin{abstract}
We show that the conditions allowing for a spontaneous breaking of the $U(1)_Y$ hypercharge gauge symmetry of the Standard Model (SM) in the early Universe are generically present in extensions of the SM addressing the generation of light neutrino masses via radiative contributions.
In such scenarios, the breaking of (hyper)charge at high-temperatures yields new possibilities for explaining the observed matter-antimatter asymmetry of the Universe. Considering for concreteness the Zee-Babu radiative neutrino mass generation model, we show that a period of hypercharge breaking prior to the electroweak phase transition could allow for successful baryogenesis via a non-conventional leptogenesis mechanism, based on the presence of charge-breaking masses for the SM leptons in the early Universe. 
\end{abstract}

\maketitle

\section{Introduction}

Contrary to Standard Model (SM) description of the  electroweak (EW) vacuum evolution from the early Universe, certain theories Beyond the SM (BSM) predict that the EW symmetry would have remained broken at high temperatures, or that the temperature of EW restoration was much higher than the EW energy scale $v = 246$ GeV~\cite{Orloff:1996yn,Gavela:1998ux,Espinosa:2004pn,Meade:2018saz,Baldes:2018nel,Glioti:2018roy,Matsedonskyi:2020mlz,Matsedonskyi:2020kuy,Biekotter:2021ysx,Carena:2021onl,Biekotter:2022kgf}. This may occur, for example, in BSM theories which feature new scalar fields $\phi_a$ charged under the EW gauge symmetry, if $\phi_a$ developed a vacuum expectation value (vev) $\left\langle \phi_a \right\rangle \neq 0$ at temperatures $T \gg 100$ GeV. In this work, we analyze an important set of such BSM scenarios, corresponding to models which address the generation of light neutrino masses via loop corrections (see e.g.~\cite{Zee:1985id,Babu:1988ki,Krauss:2002px,Ma:2006km,Aoki:2008av,Gustafsson:2012vj,Cai:2017jrq}). The presence of BSM scalar fields charged under the $SU(2)_L \times U(1)_Y$ gauge symmetry in these models is needed to achieve the breaking of the lepton number symmetry. This breaking then leads to the generation of light Majorana neutrino masses at loop level. 

We show in this Letter that radiative neutrino mass models can feature a period in the early Universe when the $U(1)_Y$ hypercharge symmetry was spontaneously broken. This would lead to an epoch during the radiation dominated era where the thermal plasma was not electrically neutral, which could have important phenomenological consequences. Among those, we identify a new mechanism to produce the observed matter-antimatter asymmetry of the Universe, from the interplay of the transient hypercharge breaking period and the interactions of the BSM scalar field(s) $\phi_a$ with the SM leptons. Together, these yield a violation of lepton number and of Charge-Parity (CP) symmetries, possibly allowing to satisfy the Sakharov conditions for baryogenesis~\cite{Sakharov:1967dj}. We analyze the conditions leading to hypercharge breaking -- and the energy range of validity of our analysis, above which the model would need to be UV completed -- and delineate the mechanism of baryogenesis for a concrete radiative neutrino mass scenario, the \textit{Zee-Babu} (ZB) model~\cite{Babu:1988ki,Cheng:1980qt}.~We nevertheless stress that these ingredients 
are common to many radiative neutrino mass models, and a broad study of the phenomenon in these models will be presented in a companion work~\cite{LLMNV2}.

%

\section{Radiative neutrino masses}

The ZB model introduces two BSM scalar fields $\kappa^+$ and $\rho^{++}$, singlets under $SU(2)_L$ and with non-zero hypercharge ($+1$ and $+2$ respectively). The tree-level scalar potential of the ZB model is  
\begin{align}
     V_{\rm tree}  = & \,\, \ov{m}^{2}_{h} |H|^{2} + \ov{m}^{2}_{\kappa} |\kappa|^{2} + \ov{m}^{2}_{\rho} |\rho|^{2} + \lambda_{h} |H|^{4}+\lambda_{\kappa} |\kappa|^{4} \nn\\ 
    & + \lambda_{\rho} |\rho|^{4} + \lambda_{\kappa \rho}|\kappa|^{2}|\rho|^{2} + \lambda_{h \kappa} |\kappa|^{2} |H|^{2}+
    \label{scapot}\\ 
    & + \lambda_{h\rho} |\rho|^{2} |H|^{2} + \left( \mu \, \rho^{++}\kappa^{-}\kappa^{-}  + \hc \right) \, .\nn
\end{align}
where $H$ is the SM Higgs $SU(2)_L$ doublet field, which in the EW broken phase and in the unitary gauge can be written as $H= (0 ,\, (v + h)/\sqrt{2})$.~The requirement that the scalar potential $V_{\rm tree} (H, \, \kappa,\, \rho)$ be bounded from below imposes that the matrix of quartic scalar couplings in Eq.~\eqref{scapot} be copositive~\cite{Kannike:2012pe}. This implies $\lambda_a \geq 0$ and \mbox{$\lambda_{ab} + \sqrt{\lambda_a \, \lambda_b} \equiv \ov{\lambda}_{ab} \geq 0$} (with $a,\,b = h, \kappa, \rho$), as well as $\sqrt{\lambda_h \lambda_{\kappa} \lambda_{\rho}} + \lambda_{h\kappa} \sqrt{\lambda_{\rho}} + \lambda_{h\rho} \sqrt{\lambda_{\kappa}} + \lambda_{\kappa\rho} \sqrt{\lambda_{h}} + \sqrt{2\,\ov{\lambda}_{h\kappa} \ov{\lambda}_{\kappa\rho} \ov{\lambda}_{h\rho}} \geq 0$.

Given the $SU(2)_L \times U(1)_Y$ transformation properties of $\kappa$ and $\rho$, the relevant Lagrangian terms in addition to $V_{\rm tree}(H, \kappa, \rho)$ read
\begin{equation}
\hspace{-2mm}    -\LL
    \supset \, \overline{L}_L \widehat{Y}_\ell\, \ell_R H +\overline{\widetilde{L}_L} f\, L_L \kappa^{+} + \overline{\ell^{c}_{R}}\, g\, \ell_{R}\, \rho^{++} + \hc
    \label{Yula}
\end{equation}
where $L_L$ and $\ell_R$ are the left-handed (LH) $SU(2)_L$ lepton doublets and right-handed (RH) charged-lepton $SU(2)_L$ singlets of the SM, respectively. We define $\widetilde{L}_{L} \equiv i \sigma_{2} L_{L}^{c}$ (with $\sigma_{2}$ the second Pauli matrix). $\widehat{Y}_\ell$ is the charged-lepton Dirac Yukawa matrix, taken to be diagonal without loss of generality. The matrices $f$ and $g$ are respectively antisymmetric and symmetric in flavour space.

\vspace{1mm}

The simultaneous presence of non-vanishing $\mu$ (from Eq.~\eqref{scapot}) and the matrices $\widehat{Y}_\ell$, $f$, $g$ yields explicit lepton number violation (LNV), and generates a Majorana mass for the SM LH neutrinos at two-loop order. The light-neutrino  mass matrix is~\cite{Babu:1988ki,Herrero-Garcia:2014hfa} $M_\nu  = \zeta\, f \,\widehat{M}_\ell\, g^\dag \, \widehat{M}_\ell \, f^T\,,$ where $\widehat{M}_\ell = \widehat{Y}_\ell\, v/\sqrt{2} = {\rm diag} (m_{e},\,m_{\mu},\,m_{\tau})$ and we have defined $\zeta\equiv I(m_\kappa^2/m_\rho^2)\times \mu/48 \pi^2 m_{\rm max}^2$, with $m_\kappa,\,m_\rho$ the masses of the BSM scalars, $m_{\rm max} = {\rm max}\, (m_\kappa,\,m_\rho)$, and $I(x)$ an $\mathcal{O}(1)$ loop integral~\cite{Herrero-Garcia:2014hfa}. To reproduce the current values of neutrino mass and mixing parameters from oscillation data~\cite{Esteban:2020cvm,Nufit} it suffices to consider non-zero $g_{\mu\mu}$, $g_{\tau\tau}$, $g_{\mu\tau}$ and $f_{jk}$ ($j,k = e,\mu,\tau$ with $j\neq k$), with $|g_{\tau\tau}|\simeq (m_{\mu}/m_{\tau}) \,|g_{\mu\tau}| \simeq (m_{\mu}/m_{\tau})^2 \, |g_{\mu\mu}|$ and \mbox{$|f_{e\mu}|\simeq|f_{e\tau}|\simeq n|f_{\mu\tau}|$} with $n \sim 0.5$ for the normal neutrino mass ordering (NO) and $n \sim 4$ for the inverse neutrino mass ordering (IO). The $g$-electron couplings ($g_{ee}$, $g_{e\mu}$, $g_{e\tau}$) play a very minor role regarding neutrino mass and mixing data. Besides, since the matrix $f$ is antisymmetric, $\det(f)=0$ and as a consequence also $\det(M_\nu)=0$, leading to the lightest neutrino being massless at the $2$-loop order.
We thus identify the absolute neutrino mass scale with the atmospheric neutrino mass difference $\sqrt{\Delta m^2_\text{atm}}\simeq 0.05$ eV~\cite{Esteban:2020cvm}, which is reproduced e.g.~for $m_{\rm max}$, $\mu$ $\sim$ TeV, $f_{jk} \sim 10^{-1}$ and $g_{\tau\tau} \sim 10^{-4}$.

The presence of lepton flavour violating (LFV) processes mediated by the BSM scalars $\kappa$ and/or $\rho$ limits the size of the entries of $g$ and $f$. For non-vanishing $f_{jk}$, $g_{\mu\mu}$, $g_{\tau\tau}$ and $g_{\mu\tau}$ the strongest limits are provided by the non-observation at 90\% C.L. of the LFV decays $\tau^-\to \mu^+\mu^-\mu^-$~\cite{Hayasaka:2010np} and $\mu\to e\gamma$~\cite{MEG:2016leq}, which lead to the respective bounds $|g_{\mu\tau}g^\ast_{\mu\mu}|<0.008\,(m_\rho/\TeV)^2$, $|f^\ast_{e\tau}f_{\mu\tau}|<0.0007\,(m_k/\TeV)^2$. The $g$-electron couplings are instead constrained dominantly by $\ell'\to3\ell$ processes. The strongest limit comes from $\mu\to3e$ searches~\cite{SINDRUM:1987nra} and leads to $|g_{e\mu}g^\ast_{ee}|<2.3\times 10^{-5}\,(m_\kappa/\TeV)^2$.
We defer a more detailed discussion to the App.~\ref{AppA} (see also Ref.~\cite{Herrero-Garcia:2014hfa}).

%
\boldmath
\section{Hypercharge breaking for \mbox{$T \gg 100$ GeV}}
\unboldmath

Let us now analyze the possibility of a $U(1)_Y$ spontaneous breaking at temperatures higher than the EW scale. To track the evolution of the vacuum during the radiation-dominated era of the early Universe we use the finite-temperature effective potential $V_{\rm eff}^T$, adding zero-temperature loop corrections as well as thermal corrections to the tree-level scalar potential from Eq.~\eqref{scapot}. These depend on the field-dependent masses $m^2_{b,f}(\phi_a)$ for bosonic and fermionic degrees of freedom (d.o.f.)~in the $\phi_a = h,\, \kappa, \, \rho$ background fields, given explicitly in the App.~\ref{AppB}. At 1-loop we have $V_{\rm eff}^T = V_{\rm tree} + V_{\rm CW} + V_{T}$, with $V_{\rm CW}$ the zero-temperature {\it Coleman-Weinberg} (CW) potential~\cite{Coleman:1973jx}, and $V_T$ containing the 1-loop thermal corrections. These are given in terms of the thermal $J$-functions for fermions ($+$) and bosons $(-)$ $J_{\pm} (x^2) = \int_0^{\infty} dy\,y^2 [1 \pm \mathrm{exp}(-\sqrt{y^2 + x^2})]$ (see e.g. Ref.~\cite{Quiros:1999jp} for a review) as
\begin{equation}
\label{thermalpartVeff}
V_{T}(\phi_a,T/T^4 
= \sum \frac{n_{b}}{2\pi^2} \, J_-\left( x_b^2\right)  - \sum \frac{n_{f}}{2\pi^2}\, J_+\left( x_f^2 \right) \, ,
\end{equation}
with $x_b = m_b(\phi_a)/T$, $x_f = m_f(\phi_a)/T$, and $n_{b}$ ($n_f$) the number of d.o.f.~of each bosonic (fermionic) species the background fields $\phi_a$ couple to, summed over in Eq.~\eqref{thermalpartVeff}. 
For $T \gg \phi_a $, we can perform a high-temperature expansion ($x^2_b,\, x^2_f \ll 1$) of the thermal $J$-functions $J_{\pm} (x^2)$, as
\begin{equation}
\begin{split}
    J_-\left(x_b^2\right) & = -\frac{\pi^4}{45} + \frac{\pi^2}{12} x_b^2 - \frac{\pi}{6}\left( x_b^2 \right) ^{3/2} + \mathcal{O}(x_b^4) \, \textrm{,} \\     
    J_+\left( x_f^2\right) & = \frac{7\pi^4}{360} - \frac{\pi^2}{24}x_f^2 + \mathcal{O}(x_f^4) \, \textrm{.}
    \end{split}
\label{thermalfunctions}
\end{equation}
The $\mathcal{O}(x^2_{b,f})$ in Eq.~\eqref{thermalfunctions} yield field-dependent $\mathcal{O}(T^2)$ contributions to $V_T$, which represent the leading 1-loop correction to $V_{\rm tree}$ at high temperature. In this limit the squared-mass term in $V_{\rm eff}^T$ for each background field $\phi_a$ reads $\ov{m}^{2}_{\phi_a} + T^2\, \mathcal{C}_{\phi_a}^0$, with $\ov{m}^{2}_{\phi_a}$ the tree-level squared-mass terms in~\eqref{scapot}, and the thermal coefficients $\mathcal{C}_{\phi_a}^0$ given by 
\begin{equation}
\begin{split}
    \mathcal{C}_{h}^0 &= \frac{\left(g'^{2} + 3 g^{2} \right)}{32} + \frac{y_{t}^{2}}{8} + \frac{\lambda_{h}}{4}  + \frac{\left( \lambda_{h\kappa} + \lambda_{h\rho} \right)}{24}\, \textrm{,}  \\
     \mathcal{C}_{\kappa}^0 &= \frac{g'^{2}}{4}  + \frac{\left(4\lambda_{\kappa} +  2\lambda_{h\kappa} + \lambda_{\kappa\rho} \right)}{12}  +  \frac{1}{24} \sum_{j \neq k  } |f_{jk}|^{2} \, \textrm{,} \\ 
    \mathcal{C}_{\rho}^0 &=  g'^{2} + \frac{\left( 4\lambda_{\rho} + 2\lambda_{h\rho} + \lambda_{\kappa\rho} \right)}{12}  + \frac{1}{24} \sum_{j,k} |g_{jk}|^{2}\, \textrm{,}
\end{split}
\label{C0fields}
\end{equation}
with $g$, $g'$ the $SU(2)_L$ and $U(1)_Y$ gauge coupling constants, and $y_t$ the top-quark Yukawa coupling (the effect of the rest of SM Yukawa couplings on $\mathcal{C}_{h}^0$ is negligible). \\

The sign of $\mathcal{C}_{\phi_a}^0$ would in principle determine whether at high temperatures $T \gg v$ the origin of field-space is destabilized ($\mathcal{C}_{\phi_a}^0 < 0$) or not ($\mathcal{C}_{\phi_a}^0 > 0$)  along the $\phi_a$ direction.~Yet, to maintain a consistent perturbative expansion in the high-temperature limit, a set of bosonic higher-loop contributions (so-called \textit{Daisy} diagrams) need to be resummed and added to the potential $V_{\rm eff}^T$ (see 
e.g. Ref.~\cite{Curtin:2016urg} for a recent discussion of this issue).~In the Arnold-Espinosa~\cite{Arnold:1992rz} prescription, the resummation of these diagrams yields the \textit{Daisy} contribution $V_{\rm D}$ added to $V_{\rm eff}^T$,
given by 
\mbox{$V_{\rm D} = T/(12 \pi) \sum_{b} \ov{n}_{b} \left[m_b^3(\phi_a) - \left( m_b^2(\phi_a) + \Pi_b \right)^{3/2} \right]$}.~To leading-order, the bosonic thermal masses $\Pi_b$ are field-independent~\cite{Carrington:1991hz}, $\Pi_b(T) = D_b T^2$, with $D_b$ constant. A high-temperature expansion of the $(m_b^2 + \Pi_b)^{3/2}$ term in the Daisy potential $V_{\rm D}$ to order $x_b^2$ yields 
\begin{equation}
\begin{split}
V_{\rm D} & \supset - \ov{n}_{b} T/(12 \pi) \times (m_b^2(\phi_i) + \Pi_b(T))^{3/2} \\
\, & = - \ov{n}_{b} T^4/(12 \pi) \times \left( D_b^{3/2} + \frac{3}{2}\, D_b^{1/2} x_b^2 + ...\right)\,.
 \end{split}
\end{equation}
These $x_b^2$ contributions result in shifts $\Delta\mathcal{C}_{\phi_a}^0$ to the thermal coefficients $\mathcal{C}_{\phi_a}^0$, i.e. $\mathcal{C}_{\phi_a} = \mathcal{C}_{\phi_a}^0 + \Delta\mathcal{C}_{\phi_a}^0$, with the shifts given by
\begin{equation}
\label{Cfields}
    \begin{split}
        \Delta\mathcal{C}_h^0
        &=
        - \frac{1}{8\pi} \left[ 6 \sqrt{2\mathcal{C}_h^0}\,\lambda_h +  \sqrt{\mathcal{C}_{\kappa}^0}\,\lambda_{h\kappa} + \sqrt{\mathcal{C}_{\rho}^0}\,\lambda_{h\rho}\right] \\
 &- \frac{1}{16\pi}\sqrt{\frac{11}{6}}g^3 -  \frac{(g^2 + g'^2)}{64\pi} \left[\sqrt{\delta_{+}} + \sqrt{\delta_{-}} \right]  \, \textrm{,} \\
  \Delta \mathcal{C}_\kappa^0
  &=
  - \frac{1}{4\pi} \left[ 
  2\sqrt{2\mathcal{C}_h^0}\,\lambda_{h\kappa} + 4 \sqrt{\mathcal{C}_{\kappa}^0}\,\lambda_{\kappa} + \sqrt{\mathcal{C}_{\rho}^0}\,\lambda_{\kappa\rho}\right] \\
 &- \frac{g'^2}{8\pi} \left[\sqrt{\delta_{+}} + \sqrt{\delta_{-}} \right]  \, \textrm{,}  \\
  \Delta\mathcal{C}_\rho^0
  &=
  - \frac{1}{4\pi} \left[ 
  2\sqrt{2\mathcal{C}_h^0}\,\lambda_{h\rho} +  \sqrt{\mathcal{C}_{\kappa}^0}\,\lambda_{\kappa\rho} + 4 \sqrt{\mathcal{C}_{\rho}^0}\,\lambda_{\rho}\right] \\
 &- \frac{g'^2}{2\pi} \left[\sqrt{\delta_{+}} + \sqrt{\delta_{-}} \right]  \, \textrm{.}
    \end{split}
\end{equation}
The coefficients $\delta_{\pm}$ above are given explicitly by $\delta_+ = 102/6 \times g'^2$, $\delta_- = 22/6 \times g^2$, related to the thermal masses for the neutral gauge bosons $Z,\, \gamma$ (their derivation is given in the App.~\ref{AppB}). 
 
\begin{figure}[h!]
\begin{centering}
\hspace{-5mm} \includegraphics[width=0.485\textwidth]{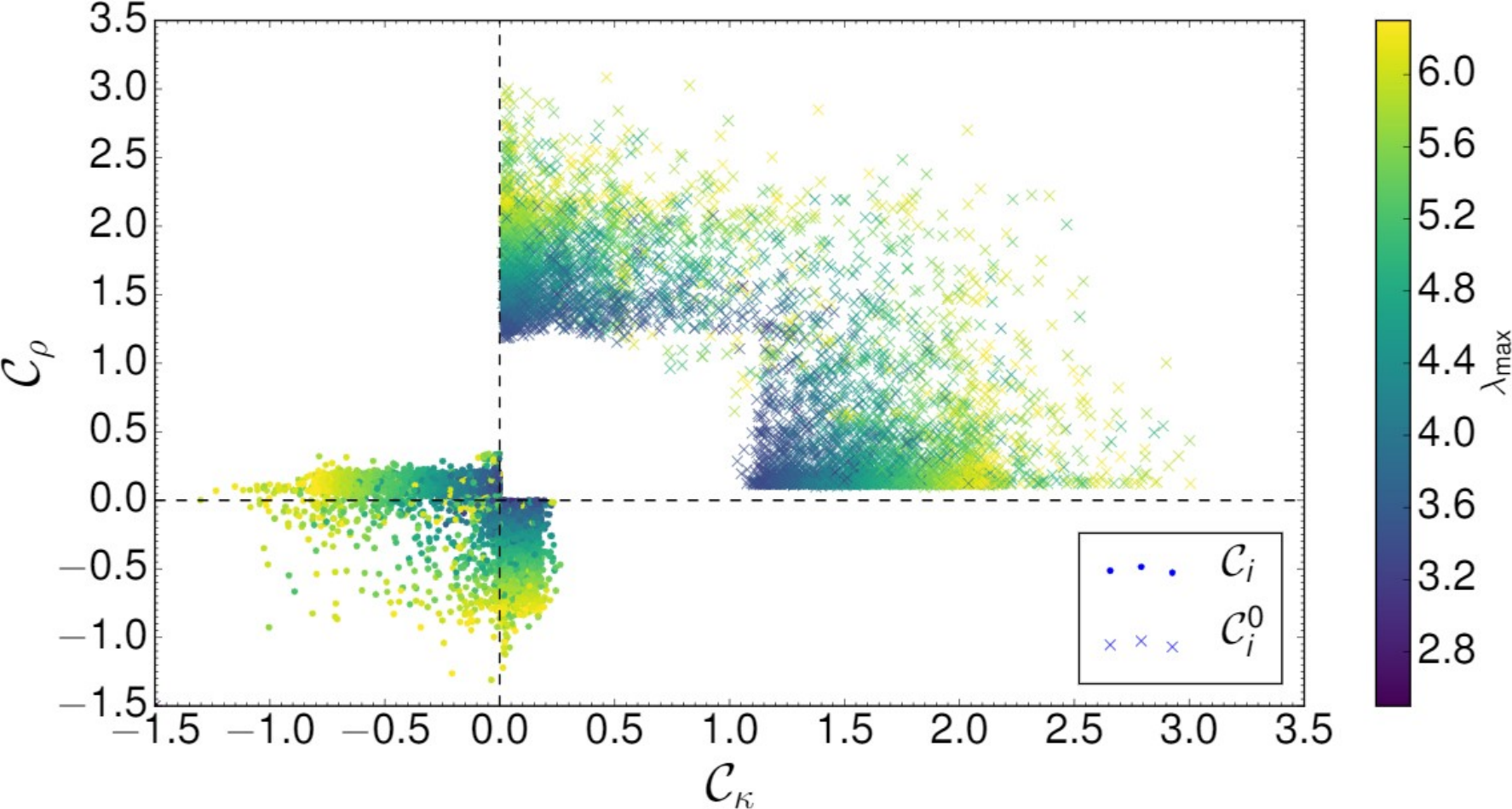} 
\hspace{1.4mm} \includegraphics[width=0.485\textwidth]{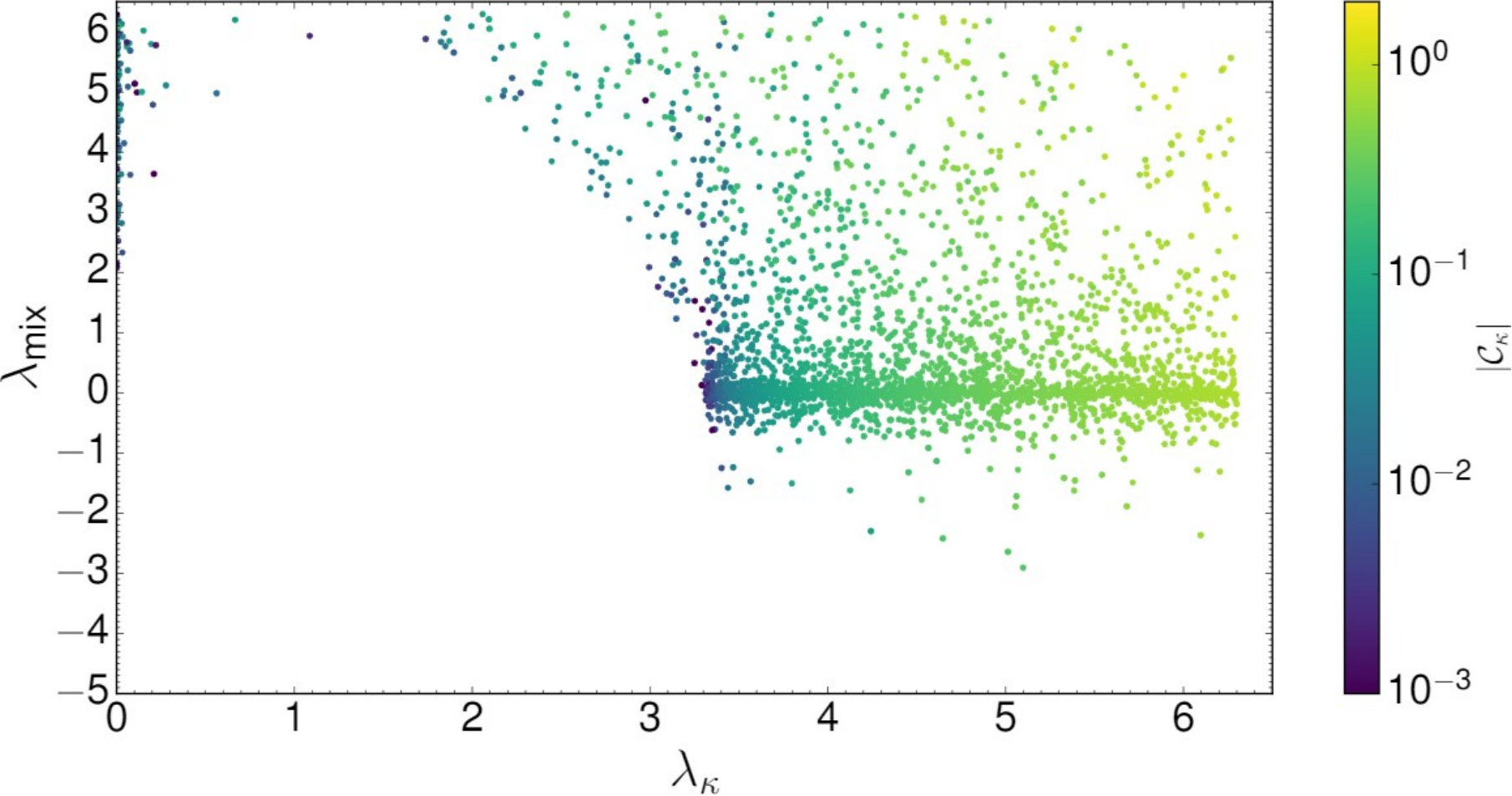} 
\caption{\em Top: $\mathcal{C}_{\kappa,\rho}$ (circles) and $\mathcal{C}^0_{\kappa,\rho}$ (crosses) values for the $\left\lbrace \lambda_{\kappa},\,\lambda_{\rho},\,\lambda_{h\kappa},\,\lambda_{h\rho},\,\lambda_{\kappa\rho} \right\rbrace$ ZB parameter scan. The colorbar represents the value of $\lambda_{\rm max} = {\rm max}\left\lbrace \lambda_i \right\rbrace$. Bottom: Parameter scan points featuring $\mathcal{C}_{\kappa} < 0$, $\mathcal{C}_{\rho} > 0$ and all $\left|\lambda_i\right| < 2\pi$. The colorbar represents the value of $\left|\mathcal{C}_{\kappa}\right|$.}
\label{fig1:Cs}
\par\end{centering}
\end{figure}

Then, the leading contributions to the finite-temperature effective potential at high temperatures are $V_{\rm eff}^T \sim V_{\rm tree}(\phi_a) + \mathcal{C}_{\phi_a} T^2 \phi_a^2$, which we refer to as the {\it Hartree} approximation~\cite{Schnitzer:1974ue}. The signs of the shifted $\mathcal{C}_{\phi_a}$ thermal coefficients consistently dictate the fate of the $SU(2)_L \times U(1)_Y$ gauge symmetry at high temperatures. For $\mathcal{C}_{\kappa} < 0$ and/or $\mathcal{C}_{\rho} <0$ the $U(1)_Y$ hypercharge symmetry will be spontaneously broken at $T \gg 100$ GeV  by non-vanishing vevs $\left\langle \kappa \right\rangle$ and/or $\left\langle \rho \right\rangle$. 
If the respective contributions from $f_{jk}$ and $g_{jk}$ to $\mathcal{C}^0_{\kappa}$ and $\mathcal{C}^0_{\rho}$ in Eq.~\eqref{C0fields} were neglected (this possibility will be justified later), the occurrence of high-temperature hypercharge breaking would solely depend on the values of the quartic couplings in Eq.~\eqref{scapot},  
$\left\lbrace \lambda_{i} \right\rbrace \equiv \left\lbrace \lambda_{\kappa},\,\lambda_{\rho},\,\lambda_{h\kappa},\,\lambda_{h\rho},\,\lambda_{\kappa\rho} \right\rbrace$.
To investigate the possibility of early Universe hypercharge breaking, we then perform a Monte Carlo scan of the $\left\lbrace \lambda_{i} \right\rbrace$ ZB parameter set, demanding boundedness-from-below on the scalar potential (as discussed previously) together with 
$\mathcal{C}_{\kappa} < 0$ and/or $\mathcal{C}_{\rho} <0$, and requiring all scanned couplings to satisfy $\left|\lambda_i\right| < 2 \pi$ to ensure perturbativity. We also focus on the ZB parameter space region with $\mathcal{C}_{h} > 0$,\footnote{The large, positive $y_t^2$ contribution to $\mathcal{C}_h^0$ generally guarantees a positive $\mathcal{C}_{h}$, except for very large ZB quartic scalar couplings.} for which the $SU(2)_L$ symmetry is restored for temperatures above the EW scale, as in the SM. In Figure~\ref{fig1:Cs} (top) we depict the scan values of $\mathcal{C}_{\kappa,\rho}$ (circles) and $\mathcal{C}^0_{\kappa,\rho}$ (crosses), showing also the maximum value of the quartic couplings $\left\lbrace \lambda_{i} \right\rbrace$ in each case, denoted by $\lambda_{\rm max}$. We observe that early Universe hypercharge breaking is only possible for $\lambda_{\rm max} > 2.5$, and that for $\lambda_{\rm max} < 2 \pi$ the coefficients $\mathcal{C}^0_{\kappa,\rho}$ are always positive, which emphasizes that the shifts $\Delta\mathcal{C}_{\phi_a}^0$ from the \textit{Daisy} contributions to $V_{\rm eff}^T$ are crucial to achieve $U(1)_Y$ spontaneous breaking in the early Universe. In Fig.~\ref{fig1:Cs} (bottom) we focus on the scenarios with $\mathcal{C}_{\rho} > 0$ and $\mathcal{C}_{\kappa} < 0$,  showing explicitly the relation between $\lambda_{\kappa}$, the quartic coupling with the largest absolute value among $\lambda_{h\kappa},\,\lambda_{h\rho},\,\lambda_{\kappa\rho}$ (denoted by $\lambda_{\rm mix}$), and $\left| \mathcal{C}_{\kappa} \right|$.

Denoting by $T_{\rm EW}$ the temperature below which the $SU(2)_L \times U(1)_Y \to U(1)_{\rm EM}$ spontaneous breaking would occur in the early Universe considering only the Higgs field direction (assuming $\mathcal{C}_{h} > 0$),\footnote{As $T_{\rm EW}$ concerns only the Higgs field direction, it does not need to correspond to the actual temperature of the EW phase transition in the ZB model if $T_{Y} < T_{\rm EW}$, see discussion below.} and $T_{Y}$ the temperature above which a vev develops for the $\rho$ and/or $\kappa$ fields in the early Universe,  we can identify two qualitatively different thermal histories:~\textit{(i)}
For $T_Y > T_{\rm EW}$ a period of $U(1)_Y$ restoration below $T_{Y}$ would precede the EW phase transition -- when the breaking of $SU(2)_L$ would be triggered in the early Universe --~\textit{(ii)} For $T_Y < T_{\rm EW}$ the hypercharge breaking epoch would last all the way down to the EW phase transition. 

%
\section{Baryogenesis}

A non-vanishing vev $\vev{\kappa}$ and/or $\vev{\rho}$ in the early Universe would induce charge-breaking mass terms for the SM leptons from 
Eq.~\eqref{Yula},~i.e. $\ov{\nu^c_L}\ell_L + {\rm h.c.}$ and $\overline{\ell^{c}_R} \ell_R$ for $\vev{\kappa} \equiv v_{\kappa} \neq 0$ and $\vev{\rho} \equiv v_{\rho} \neq 0$, respectively. This would lead to lepton mass eigenstates $\ell_{M_i}$ stemming from these mass matrices (e.g.~for $v_{\rho} \neq 0$, $v_{\kappa} = 0$, the $\ell_{M_i}$ are Majorana-like eigenstates of the $g$ flavour matrix in Eq.~\eqref{Yula}, with $m_{\ell_{M_i}} \propto v_{\rho}$). The Majorana nature of such mass terms results in LNV, which 
combined with the CP violation from the complex phases of the $g$ and $f$ flavour matrices, could yield successful baryogenesis, as we now discuss.\\

Let us first consider the case $T_Y > T_{\rm EW}$, where a period of hypercharge restoration for $T \in [T_{\rm EW},\,T_Y]$ precedes the EW phase transition:~for $T > T_Y$, during the hypercharge-breaking epoch, the $\ell_{M_i}$ states 
would have LNV decays, e.g.~both $\ell_{M_i} \to \overline{\ell}_{M_j} \ell_{M_k} \ell_{M_l}$ and $\ell_{M_i} \to \ell_{M_j} \ell_{M_k} \ell_{M_l}$. 
Such decays, if occurring out-of-equilibrium and with CP violation from the interference between tree-level and 1-loop decay amplitudes, could generate a net lepton asymmetry. This would then be converted into a baryon asymmetry via SM sphaleron processes prior to the EW phase transition.  The overall physical mechanism would be analogous to leptogenesis~\cite{Fukugita:1986hr} in the high-scale Type-I Seesaw mechanism for neutrino mass generation (see also Refs.~\cite{Buchmuller:2004nz,Davidson:2008bu,Biondini:2017rpb} for leptogenesis reviews), with the $\ell_{M_i}$ states (of Majorana nature) playing the role of the RH neutrinos. Yet, this leptogenesis mechanism faces several important challenges: {\it(i)} Since for 
$T \to T_Y$ the vevs $v_{\kappa}, \, v_{\rho}$ must vanish, $m_{\ell_{M_i}}/T \ll 1$ as $T$ approaches $T_Y$, and the LNV decays of $\ell_{M_i}$ will then inevitably occur in thermal equilibrium.~{\it(ii)} For $T \gg T_Y$ the vevs $v_{\rho}$ and/or $v_{\kappa}$ grow as $\propto T$, so it may not be possible to achieve out-of-equilibrium decays of $\ell_{M_i}$ also in that limit. {\it(iii)} The presence of CP violation in the interference between tree-level and loop decay amplitudes of $\ell_{M_i}$ require off-diagonal $g_{jk}$ and/or $f_{jk}$; while this is indeed needed to fit neutrino oscillation data in the ZB model, it will also inevitably connect the decays and inverse decays of the various $\ell_{M_i}$ eigenstates, which would in principle preclude the possibility of secluding a lepton asymmetry in a specific flavour. Overall, we do not attempt to quantify the possibility of achieving leptogenesis in this way, but point out that it seems challenging and that flavour effects would play a key role in such case.\\

For $T_{\rm EW} > T_{Y}$ the hypercharge-breaking period would extend all the way down to the EW phase transition. This provides a different way in which successful baryogenesis could be achieved. For $T \in [T_Y,\, T_{\rm EW}]$ the hypercharge breaking vacuum and the EW vacuum would coexist, becoming degenerate at a critical temperature $T_c$ (with $T_Y < T_c < T_{\rm EW}$). A first-order phase transition would occur from the former vacuum to the latter below $T_c$, and this transition would as such proceed by nucleation and growth of true vacuum bubbles, of the EW phase. The space-time dependent $\vev{\rho}$, $\vev{\kappa}$ background of the bubble boundaries/walls would induce varying complex masses (yielding CP violation) for the SM leptons across the walls, which could lead to a lepton asymmetry as well as to a baryon asymmetry through EW sphalerons. The mechanism would bear many similarities with EW baryogenesis~\cite{Cohen:1993nk,Trodden:1998ym,Morrissey:2012db,Konstandin:2013caa}, with the relevant CP violating processes linked to the leptons rather than the SM quarks (see e.g. Ref.~\cite{Pascoli:2016gkf} for a related setup). We here focus on this, more promising possibility.\footnote{Since at the EW phase transition the vevs $\vev{\rho}$ and/or $\vev{\kappa}$ shut-off abruptly, this could also allow the LNV decays of the $\ell_{M_i}$ states as $T \to T_{\rm EW}$ to occur out-of-equilibrium, as required for the baryogenesis mechanism outlined previously.} \\

Given a specific $\left\lbrace \lambda_{i} \right\rbrace$ set (with negative $\mathcal{C}_{\kappa}$ and/or $\mathcal{C}_{\rho}$), the values of the BSM scalar masses $m_\rho$ and $m_\kappa$ determine the details of the thermal history. 
The {\it Hartree} approximation for $V_{\rm eff}^T$ is particularly convenient to gain analytical insight on the respective parameter space regions. We now analyze the case $\mathcal{C}_{\rho} < 0$, $\mathcal{C}_{\kappa} > 0$ and leave that of $\mathcal{C}_{\kappa} < 0$, $\mathcal{C}_{\rho} > 0$ for App.~\ref{AppC}.  

We first require that the Universe evolve from a high-temperature $(\vev{h}, \vev{\kappa},\vev{\rho}) = (0,0,v_{\rho})$ minimum, which sets the conditions $\mathcal{C}_{\kappa} + \lambda_{\kappa\rho}\left|C_{\rho}\right|/(2\lambda_{\rho}) > 0$ and $\mathcal{C}_{h} + \lambda_{h\rho} \left|C_{\rho}\right|/(4\lambda_{\rho}) > 0$. Then, to achieve $T_{\rm EW} - T_{Y} \equiv \Delta T > 0$,w this high-$T$ minimum will need to remain stable for $T < T_{\rm EW}$, and become degenerate with the EW minimum $(\vev{h}, \vev{\kappa},\vev{\rho}) = (v_{h},0,0)$ at $T_c$ (with $v_{h}(T = 0) \equiv v$). The maximum $m_\rho$ mass compatible with $\Delta T > 0$ is $m_{\rho}^{\rm max} = \sqrt{(\left|C_{\rho}\right|/4\, C_h)\, m_h^2 + (\lambda_{h\rho}/2) \, v^2}$. At the same time, as $m_{\rho}$ decreases the system reaches $T_{Y} \to 0$, with the hypercharge-breaking vacuum $(0,0,v_{\rho})$ being metastable at $T = 0$. This yields a lower bound $m_{\rho}^{\rm min}$ (for $T_c \to 0$) below which the EW vacuum is no longer the absolute minimum of the system at $T= 0$. In  Fig.~\ref{fig2:point1} we show the ($m_\kappa$, $m_\rho$) region yielding the desired thermal history (with $\Delta T > 0$) for a specific benchmark $\left\lbrace \lambda_{i} \right\rbrace$ -- BM1 with $ \lambda_{\kappa} = 0.118$, $\lambda_{\rho} = 5.44$, $\lambda_{h\kappa}=4.70$, $\lambda_{h\rho}=-0.097$, $\lambda_{\kappa\rho}=-0.052$, $C_h=0.042$ -- with $\mathcal{C}_{\rho} = -0.85 < 0$ and $\mathcal{C}_{\kappa} = 0.048 > 0$.  We depict the value of $v_h(T_c)/T_c$ in the relevant ($m_\kappa$, $m_\rho$) region, noting that $v_h(T_c)/T_c > 1$ yields a first-order EW phase transition strong enough for successful baryogenesis~\cite{Quiros:1999jp}.  We note that within the whole viable region we have $T_c < T_{\rm EW} \simeq  300$ GeV.
We also show in Fig.~\ref{fig2:point1} the current bounds on the masses $m_{\kappa}$, $m_{\rho}$ from LHC and LEP searches for the ZB scalars $\kappa$ and $\rho$: {\it (i)} Existing LHC searches for right-handed sleptons (supersymmetric partners of the SM RH charged-leptons) decaying into a SM lepton $\ell = e, \mu$~\cite{ATLAS:2014zve,ATLAS:2019lff,ATLAS:2022hbt,CMS:2020bfa} or $\ell = \tau$~\cite{ATLAS:2019gti,CMS:2022syk}, and a massless neutralino, constrain the properties of $\kappa$ (gauge quantum numbers and decay modes are equal), excluding masses in the approximate range $m_{\kappa}\in [150, 250]$ GeV for NO and $m_{\kappa}\in [130, 300]$ GeV for IO at 95\% C.L. (these are derived in the App.~\ref{AppD}). {\it (ii)} LEP searches for sleptons yield the limit $m_{\kappa} \gtrsim 100$ GeV~\cite{LEPsleptons}, corresponding to the approximate pair production reach of LEP. {\it (iii)} Current LHC searches for the doubly charged scalar $\rho$~\cite{ATLAS:2022pbd} in the di-leptonic decay mode $\rho^{\pm\pm} \to \ell_j^{\pm} \ell_k^{\pm}$ put a 95\% C.L. bound $m_{\rho} > 900$ GeV assuming that all flavour combinations $j,\, k = e,\, \mu,\,\tau$ have the same branching fraction, and no other decay modes exist. These limits are nevertheless evaded when the decay $\rho^{\pm\pm} \to \kappa^{\pm} \kappa^{\pm}$ is open and dominates (which is generally the case for the values of $g_{jk}$ and $\mu$ considered in this work). Thus, we must require $m_{\rho} > 2 \, m_{\kappa}$ to evade current LHC bounds on the existence of $\rho$. \\

All the above results, derived in the {\it Hartree} approximation, are robust once the full 1-loop thermal potential $V_T$ and the full \textit{Daisy} contribution $V_{\rm D}$ are included, as shown in the App.~\ref{AppE}. Besides, the (zero-temperature) 1-loop corrections to the $\{\lambda_i\}$, e.g. $\delta \lambda_{h\rho} \sim  \lambda_{h\kappa}/16\pi^2 \times (\mu/m_\rho)^2$, $\delta \lambda_{\kappa\rho} \sim {\rm max} (\lambda_\kappa,\lambda_\rho)/16\pi^2 \times (\mu/m_\rho)^2$, are small compared to the corresponding $\{\lambda_i\}$ tree-level values. Finally, the values of $C_\rho$ and $C_\kappa$ for both benchmarks justify neglecting the Yukawa contributions from $g_{jk}$ and $f_{jk}$ to Eq.~\eqref{C0fields}, which are smaller than $10^{-3}$ for the Yukawa values that fit neutrino oscillation data.

\begin{figure}[h!]
\begin{centering}
\includegraphics[width=0.495\textwidth]{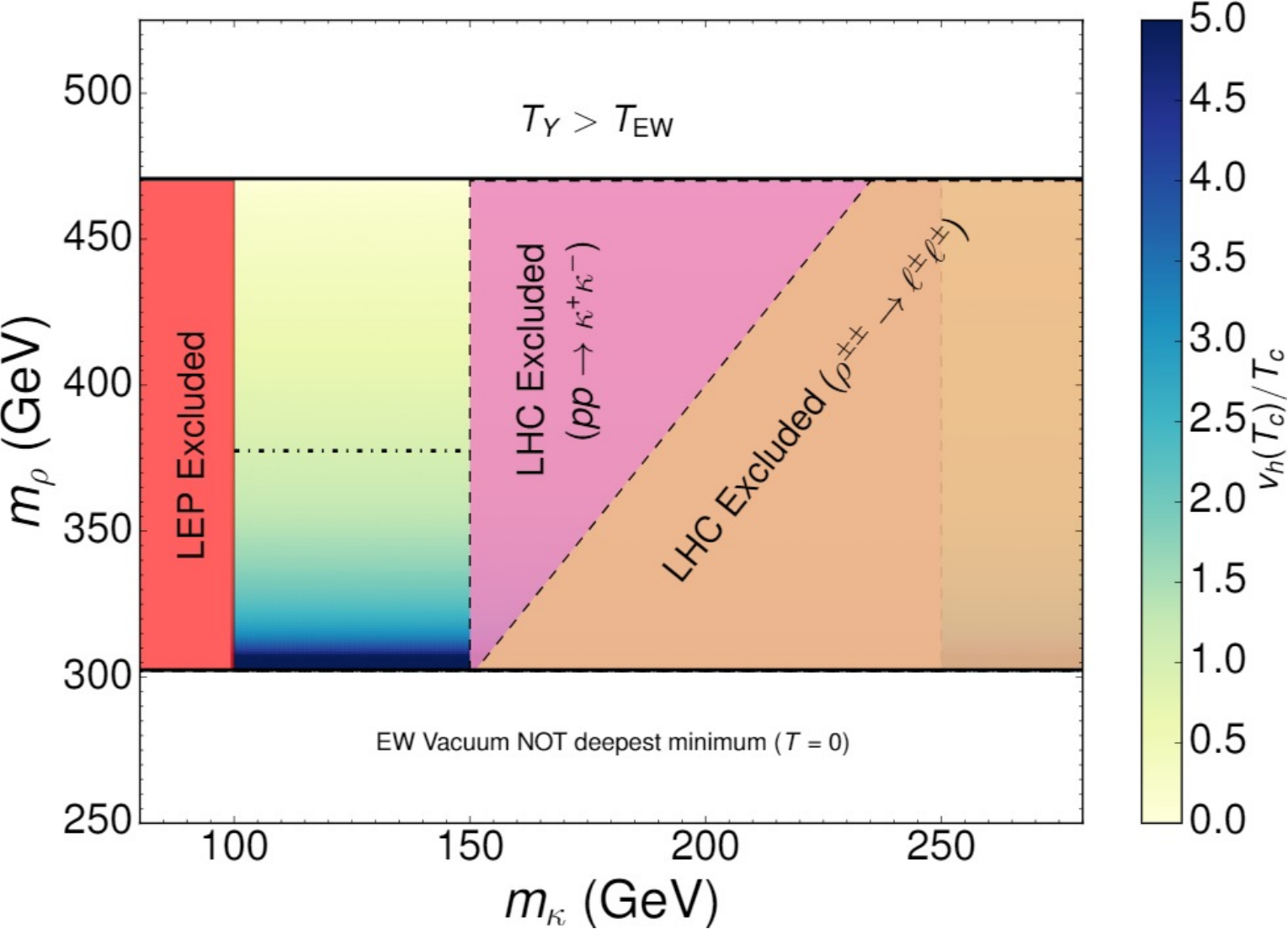} 
\caption{\em Viable ($m_{\kappa}, m_{\rho}$) parameter space of baryogenesis (corresponding to $T_{\rm EW} - T_{Y} > 0$) for benchmark scenario BM1 with $\mathcal{C}_{\rho} < 0$, $\mathcal{C}_{\kappa} > 0$ (see main text). The colorbar indicates the strength of the phase transition $v_h(T_c)/T_c$, with the dash-dotted line corresponding to $v_h(T_c)/T_c = 1$. The red region is excluded by LEP searches~\cite{LEPsleptons}, while the magenta and orange regions are excluded by LHC searches for $p p \to \kappa^+ \kappa^-$~\cite{CMS:2020bfa} and
$p p \to \rho^{++} \rho^{--}$~\cite{ATLAS:2022pbd}, respectively. White areas correspond to theoretical bounds: either the EW vacuum is not the deepest minimum at $T=0$ or $T_Y>T_{\rm EW}$ that would prevent the coexistence of hypercharge and EW breaking vacua. 
}
\label{fig2:point1}
\par\end{centering}

\end{figure} 

%
\section{Additional Comments}
We note that setting $\mu=0$ in the ZB model switches-off the explicit LNV source, and no neutrino masses are generated at $T = 0$ since lepton number is conserved (with $\rho$ and $\kappa$ necessarily charged under it). Yet, at $T\neq0$ the singlet scalar vev(s) which cause hypercharge breaking also break lepton number spontaneously. Baryogenesis could then still occur for $\mu=0$. In this case the $g$ and $f$ couplings are not constrained to reproduce the neutrino masses and PMNS mixing values, yet neutrino masses and mixings would remain unexplained.

We also stress that in the ZB model, the breaking of hypercharge at temperatures above the EW scale generally relies on at least one of the $\{\lambda_i\}$ couplings being sizable -- recall BM1 -- . It will thus become non-perturbative not far above the TeV scale, point at which the ZB model would need to be UV completed --  this may not occur for other radiative neutrino mass models, an issue we will explore elsewhere~\cite{LLMNV2} -- . The renormalization group evolution (RGE) of the $\{\lambda_i\}$ set (see~\cite{Herrero-Garcia:2014hfa}) is then needed to ascertain the range of validity of the theory. Using the RGE equations for the ZB model from~\cite{Herrero-Garcia:2014hfa}, we compute the energy scale $\Lambda_{\rm NP}$ for which the largest coupling within the $\{\lambda_i\}$ set reaches the perturbativity limit according to the naive dimensional analysis~\cite{Manohar:1983md,Gavela:2016bzc}. For BM1 we find $\Lambda_{\rm NP}$ slightly above the TeV scale, such that there is a sizable (although not vastly large) temperature range where hypercharge is indeed broken (recall that $T_c < 300$ GeV for BM1 and the ZB does not need a UV completion).

\acknowledgments
The authors are grateful to Jos\'e Ram\'on Espinosa, Enrique Fern\'andez-Mart\'inez, Miguel Escudero and Arturo de Giorgi for useful discussions and comments, as well as to Kaladi Babu for correspondence in relation to the idea behind this work. J.M.N. thanks Thomas Biekotter and Christoph Borschensky for help in analyzing the limits from Ref.~\cite{CMS:2020bfa}.~The work of J.M.N. was supported by the Ram\'on y Cajal Fellowship contract RYC-2017-22986.
The authors acknowledge partial financial support by the Spanish Research Agency (Agencia Estatal de Investigaci\'on) through the grant IFT Centro de Excelencia Severo Ochoa No CEX2020-001007-S and by the grants PID2019-108892RB-I00 and PID2021-124704NB-I00 funded by MCIN/AEI/10.13039/501100011033, by the European Union's Horizon 2020 research and innovation programme under the Marie Sk\l odowska-Curie grant agreements No 860881-HIDDeN and 101086085-ASYMMETRY.

\appendix

\section{Neutrino Mixing and LFV Constraints}
\label{AppA}
As discussed in the main text, neutrinos acquire Majorana masses at two-loop level in the ZB model, via the Feynman diagram
\begin{equation*}
    \begin{tikzpicture}
     \begin{feynman}
     \vertex (a0);
     \vertex at ($(1cm,0)+(0,0)$) (a1);
    \vertex at ($(2cm,0)+(0,0)$) (b1);
     \vertex at ($(3cm,0)+(0,0)$) (c1);
     \vertex at ($(-1cm,0)+(0,0)$) (a2);
     \vertex at ($(-2cm,0)+(0,0)$) (b2);
     \vertex at ($(-3cm,0)+(0,0)$) (c2);
     \vertex at ($(-1cm,0)+(0,-0.7cm)$) (d1);
     \vertex at ($(1cm,0)+(0,-0.7cm)$) (d2);
     \vertex at ($(0,0)+(0,2cm)$) (d3);
     \diagram*{
     (c2) -- [fermion, edge label=\(\nu_{L}\)] (b2),
     (b2) -- [anti fermion, edge label=\(\ell_{L}\)] (a2),
     (a2) -- [anti fermion, edge label=\(\ell_{R}\)] (a0),
     (c1) -- [fermion, edge label=\(\nu_{L}\)] (b1),
     (a1) -- [fermion, edge label=\(\ell_{L}\)] (b1),
     (a0) -- [fermion, edge label=\(\ell_{R}\)] (a1),
     (d3) -- [scalar, edge label=\(\rho^{--}\)] (a0),
     (a2) -- [scalar, insertion= 1] (d1),
     (a1) -- [scalar, insertion= 1, edge label] (d2),
     (b2) -- [scalar,out=90, in=180, , edge label=\(\kappa^{-}\)] (d3),
     (d3) -- [scalar,out=0, in=90, edge label=\(\kappa^{-}\)] (b1),
     };
     \end{feynman}
    \end{tikzpicture}
\end{equation*}
with $\times$ being SM charged-lepton mass insertions. The Majorana neutrino mass matrix 
$M_\nu = \zeta\, f \,\widehat{M}_\ell\, g^\dag \, \widehat{M}_\ell \, f^T$ can be be diagonalized by a unitary transformation
\be
\widehat M_\nu=U\, M_\nu\,U^T\,,
\ee
being $U$ the Pontecorvo-Maki-Nakawaga-Sakata (PMNS) lepton mixing matrix. As discussed in~\cite{Herrero-Garcia:2014hfa, Schmidt:2014zoa}, reproducing the measured neutrino oscillation data imposes a specific pattern on the entries of the flavour matrices $g$ and $f$.  First, for NO one has the relation $g^\ast_{\tau\tau}m_\tau^2\simeq g^\ast_{\mu\tau}m_\mu m_\tau\simeq
g^\ast_{\mu\mu}m_\mu^2$ for nearly maximal $\theta_{23}$ (see e.g.~Eq.~(30) of~\cite{Schmidt:2014zoa}), leading to a clear hierarchy among the entries of the matrix $g$: $|g_{\tau\tau}| \sim 0.06\, |g_{\mu\tau}| \sim 0.003 \,|g_{\mu\mu}|$. In addition, one obtains the relation $\zeta\,f_{\mu\tau}^2\left|m_a\,m_b\,g^\ast_{ab}\right| \simeq 
\sqrt{\Delta m^2_\text{atm}} \,{\rm cos}^2 (\theta_{23}) \sim 0.025\text{ eV}$, with $a,b=\mu,\tau$.  
For IO, the values of the CP phases in the PMNS play a deeper role and the relative weights between the entries of $g$ may vary, yet a certain (albeit milder than for NO) hierarchy $|g_{\tau\tau}| \ll \,|g_{\mu\tau}| \ll |g_{\mu\mu}|$ is also granted. We remark that the remaining entries of $g$ only play a marginal role in fitting neutrino oscillation data, and can be taken as vanishing for that purpose. At the same time, the vanishing of det($f$) -- due to the antisymmetric nature of $f$ -- implies correlations between the $f_{jk}$, such that $f_{e\mu}$ and $f_{e\tau}$ can be obtained in terms of $f_{\mu\tau}$ and the neutrino oscillation data. In particular, for NO we have $f_{e\mu}\sim f_{e\tau}\sim f_{\mu\tau}/2$, whereas for IO we have $f_{e\mu}\sim-f_{e\tau}$ and $|f_{e\mu}|\sim |f_{e\tau}|\sim 4|f_{\mu\tau}|$.  \\

Additional constraints on the entries of the flavour matrices $g$ and $f$ arise from LFV processes mediated by $\kappa$ and/or $\rho$. Considering $g_{\mu\mu}$, $g_{\tau\tau}$ and $g_{\mu\tau}$ as the only non-vanishing entries of $g$ (in agreement with neutrino oscillation data, as discussed above), the strongest limits are provided by the $\mu\to e\gamma$~\cite{MEG:2016leq}, $\tau^-\to \mu^+\mu^-\mu^-$~\cite{Hayasaka:2010np} and $\tau\to \mu\gamma$~\cite{Belle:2021ysv} decay searches (the complete list of constraining processes can e.g.~be found in~\cite{Herrero-Garcia:2014hfa}). The respective $90\%$ C.L. bounds on the $f_{jk}$ and $g_{jk}$ entries read
\begin{align}
&|f^\ast_{e\tau}f_{\mu\tau}|<0.0007\,(m_\kappa/\TeV)^2\nn\\
&|g_{\mu\tau}g^\ast_{\mu\mu}|<0.008\,(m_\rho/\TeV)^2\\
&|f^\ast_{e\mu}f_{e\tau}|^2\dfrac{m^4_\rho}{m^4_\kappa}+16|g^\ast_{\mu\mu}g_{\mu\tau}+g^\ast_{\mu\tau}g_{\tau\tau}|^2<0.67\,(m_\rho/\TeV)^4\,.\nn
\end{align}
We stress that for non-vanishing $g_{ee}$, $g_{e\mu}$, $g_{e\tau}$ entries, the above bounds on $f_{jk}$ from 
$\mu\to e\gamma$ would be modified. At the same time, these entries would be bounded by other LFV processes, the strongest bound coming from $\mu^-\to e^+e^-e^-$ decay searches~\cite{SINDRUM:1987nra}. In this sense, we remark that the ZB model can only accommodate the observed neutrino oscillation parameters within $3\sigma$ \cite{Irie:2021obo}.

\section{Field-dependent masses \& thermal masses in the Zee-Babu model}
\label{AppB}
The zero-temperature field-dependent masses $m_{b,f}(\phi_a)$ for the various bosonic and fermionic d.o.f. of the ZB model in $h$, $\kappa$, $\rho$ background fields are given by
\begin{align}
    m_h^2(h,\kappa,\rho) &= \ov{m}^{2}_h + 3\lambda_h h^2 + \lambda_{h\kappa}\,\kappa^2 + \lambda_{h\rho} \,\rho^2\\
    m_{\chi}^2(h,\kappa,\rho) &= \ov{m}^{2}_h + \lambda_h h^2 + \lambda_{h\kappa}\,\kappa^2 + \lambda_{h\rho} \,\rho^2 \\
    m_{\kappa}^2(h,\kappa,\rho) &= \ov{m}^{2}_{\kappa} + \frac{1}{2}\lambda_{h\kappa}h^2 + 4\lambda_{\kappa}\,\kappa^2 + \lambda_{\kappa\rho}\, \rho^2\\
    m_{\rho}^2(h,\kappa,\rho) &= \ov{m}^{2}_{\rho} + \frac{1}{2}\lambda_{h\rho} h^2 + \lambda_{\kappa\rho}\, \kappa^2 + 4\lambda_{\rho}\, \rho^2\\   
    m_{W}^2(h) &= \frac{g^2}{4}h^2 \\
        m_{t}^2(h) &= \frac{m_{t}^2}{v^2} h^2 \\
        m_{\ell_i,L}^2(\kappa) &\simeq \sum_{j \neq i} |f_{ij}|^{2} \kappa^2 \\
    m_{\ell_i,R}^2(\rho) &\simeq \sum_{j} |g_{ij}|^{2} \rho^2
\end{align}
with some obvious abuse of notation in the background-filed labelling. $\chi$ are the Goldstone d.o.f. in the SM Higgs doublet and $m_{\ell_i,L}^2$, $m_{\ell_i,R}^2$ the field-dependent masses for the SM charged leptons (with the Higgs contribution neglected). Moreover, we only include the top quark contribution (the remaining quarks are neglected as their Yukawas are very small).
The respective number of d.o.f. for each species is $n_h = 1$, $n_{\chi} = 3$, $n_{\kappa} = n_{\rho} = 2$, $n_W = 6$, $n_t = 12$, $n_{\ell_i}^{L,R} = 4$. 

The thermal (Debye) masses $\Pi_b(T)$ for the bosonic d.o.f. are given at leading-order by 
\begin{align}
    \Pi_h &= \Pi_{\chi} = 2\, \mathcal{C}_{h}^0 \, T^2 \\
    \Pi_{\kappa} &= \mathcal{C}_{\kappa}^0\, T^2 \\
    \Pi_{\rho} &= \mathcal{C}_{\rho}^0\, T^2 \\
    \Pi_W^{\textrm{L}} &= \frac{11}{6}g^2\, T^2 \\
    \Pi_B^{\textrm{L}} &= \frac{51}{6}g'^2\, T^2 
\end{align}
with $\Pi_{W,B}^{\textrm{L}} $ the thermal masses for the longitudinal $SU(2)_L$ ($W$) and $U(1)_Y$ ($B$) gauge bosons (see Refs.~\cite{Carrington:1991hz,Kapusta:2007xjq} for theoretical details and e.g.~\cite{Breitbach:2018kma,Comelli:1996vm} for exemplary non--SM applications). The corresponding thermal masses for the transverse gauge bosons vanish at leading order, i.e. $\Pi_{W,B}^{\textrm{T}} = 0$.

We have so far left aside the field-dependent masses of the $Z$ and the $\gamma$ gauge bosons. For their longitudinal components, the squared-masses are given (including the effect of the thermal masses $\Pi_W^{\textrm{L}}$ and $\Pi_B^{\textrm{L}}$) by the eigenvalues of the matrix
\begin{align}
\label{matrixZgamma}
\begin{pmatrix}
\frac{g^2 h^2}{4} + 2\,\Pi^{\textrm{L}}_{W} & -\frac{gg' h^2}{4}\\
-\frac{gg' h^2}{4} & 2\,g'^2(\frac{h^2}{8} + \kappa^2 + 4\rho^2) + 2\,\Pi^{\textrm{L}}_{B}
\end{pmatrix} \, .
\end{align}
For their transverse components the situation is analogous, but with no-thermal masses at leading order.
The eigenvalues of Eq.~\eqref{matrixZgamma} are given by
\begin{equation}
\label{fullthermalmasses}
\begin{split}
m_{Z,\gamma}^2(h,\kappa,\rho) =& \frac{1}{8} \left[ 8 (\Pi^{\textrm{L}}_{W} + \Pi^{\textrm{L}}_{B} ) + (g^2 + g'^2) h ^2 \right.\\
 &+ \left. 8\, g'^2 (\kappa^2  + 4\, \rho^2) \pm \Delta \right]
\end{split}
\end{equation}
where $\Delta$ is given by
\begin{eqnarray}
\label{discriminant}
\Delta^2 &= &\left[(g^2 + g'^2) h^2 + 8 g'^2\left(\kappa^2 +4  \rho^2\right)+ 8 (\Pi^{\textrm{L}}_{W}+
\Pi^{\textrm{L}}_{B})\right]^2 \nonumber\\
&&-32 \left( g^2 h^2 + 8 \Pi^{\textrm{L}}_{W} \right) \left[ g'^2\left(\kappa^2 +4  \rho^2\right) + \Pi^{\textrm{L}}_{B} \right]\\
&& -32 \,\Pi^{\textrm{L}}_{W} \, g'^2 h^2 \, . \nonumber
\end{eqnarray}
We note that $\Delta$ does not contribute to $V_T$ at order $\mathcal{O}(T^2)$, since for the longitudinal components of $Z$ and $\gamma$ we have $n_Z = n_{\gamma} = 1$ and $\Delta$ cancels out in the sum over d.o.f.\ in Eq.~(\color{magenta} 3\color{black}) of the main text. Yet, it does contribute to $V_{\rm eff}^T$ beyond this order. 
An expansion of $\Delta$ in powers of $1/T^2$ (thus only valid for high-temperatures) yields
\begin{eqnarray}
\label{deltaexpansion}
\Delta &=& 8\sqrt{(a-b)^2}\, T^2 \\
&&- \frac{\left|a-b\right|\left[8 g'^2 (\kappa^2 + 4 \rho^2) - (g^2 - g'^2) h^2\right]}{a-b} \nonumber\\
&& + \mathcal{O}\left(T^{-2}\right) \nonumber
\end{eqnarray}
with $a=11g^2/6$, $b=51g'^2/6$.
Retaining only the leading order terms, $\mathcal{O}(T^2)$, in $\Pi^{\textrm{L}}_{W}$, $\Pi^{\textrm{L}}_{B}$ and $\Delta$, the field-dependent masses for $Z$ and $\gamma$ finally read
\begin{equation}
\label{fullthermalmasse3}
\begin{split}
m_{Z,\gamma}^2(h,\kappa,\rho) =  & \left[\frac{(g^2 + g'^2) h ^2}{8} + g'^2 (\kappa^2  + 4\, \rho^2)+ \Pi_{Z,\gamma} \right] \, .
\end{split}
\end{equation}
with
\begin{eqnarray}
    \Pi(T)_{Z,\gamma} &=& \Pi^{\textrm{L}}_{W}(T) + \Pi^{\textrm{L}}_{B}(T) \pm \sqrt{(a-b)^2}T^2 \\
    &=& \left( \frac{11}{6}g^2 + \frac{51}{6}g'^{2} \pm \frac{1}{6} |11g^2-51g'^{2}| \right)T^2\nn \\
    &\equiv& \delta_{\pm}T^2 \, .\nn
\end{eqnarray}

\begin{figure}[h!]
\begin{centering}
\includegraphics[width=0.495\textwidth]{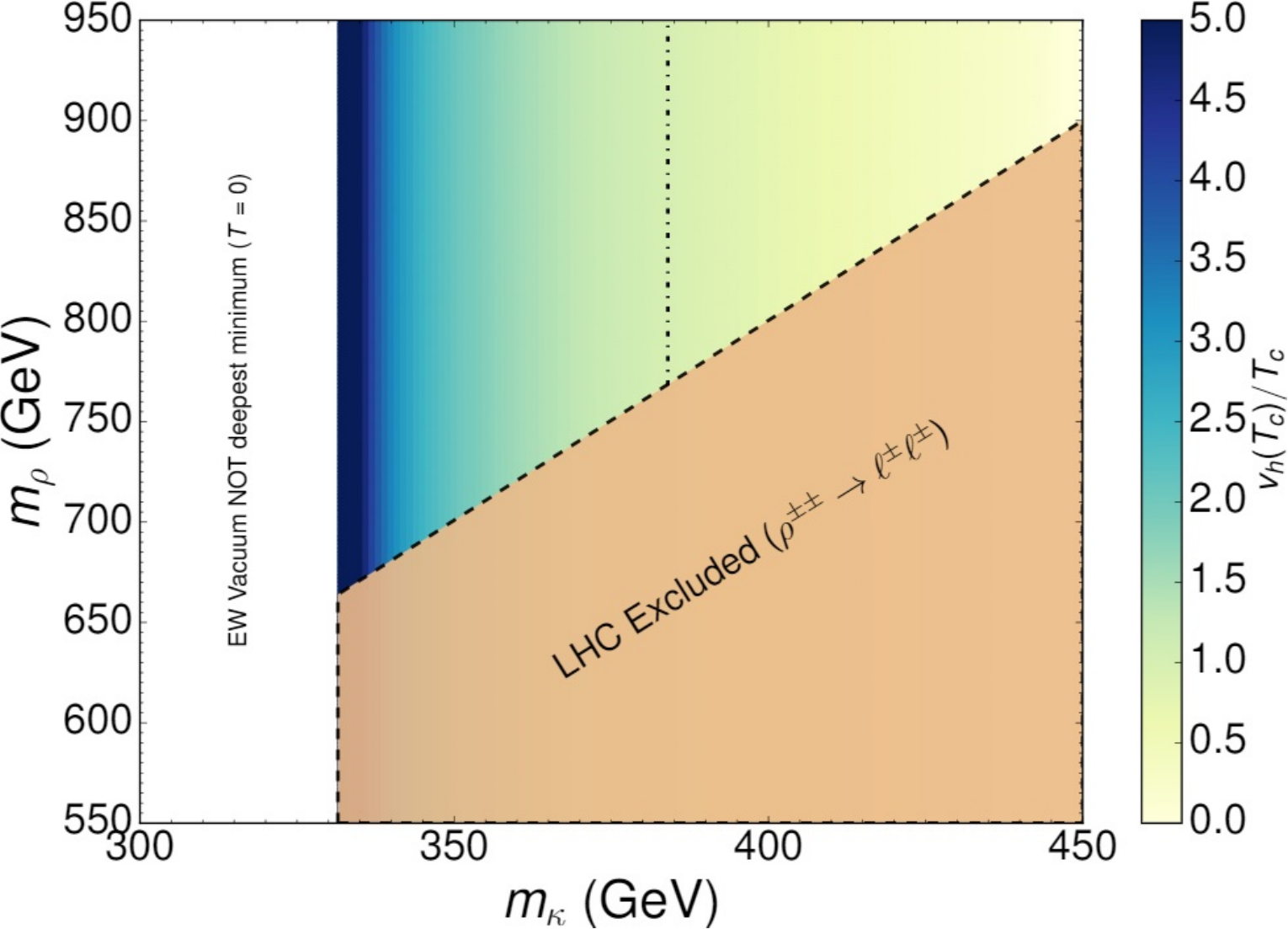} 
\caption{\em Viable ($m_{\kappa}, m_{\rho}$) parameter space of baryogenesis (corresponding to $T_{\rm EW} - T_{Y} > 0$) for benchmark scenario BM2 with $\mathcal{C}_{\rho} > 0$, $\mathcal{C}_{\kappa} < 0$ from Tab.~\ref{tab1}. The colorbar indicates the strength of the phase transition $v_h(T_c)/T_c$, with the dash-dotted line corresponding to $v_h(T_c)/T_c = 1$. The orange region is excluded by LHC searches for 
$p p \to \rho^{++} \rho^{--}$~\cite{ATLAS:2022pbd}. White areas correspond to theoretical bounds: either the EW vacuum is not the deepest minimum at $T=0$ or $T_Y>T_{\rm EW}$ that would prevent the coexistence of hypercharge and EW breaking vacua. 
}
\label{fig2:point1}
\par\end{centering}
\end{figure} 

\boldmath
\section{Baryogenesis analysis for the $\mathcal{C}_{\rho} > 0$, $\mathcal{C}_{\kappa} < 0$ setup}
\label{AppC}
\unboldmath
The baryogenesis setup with $\mathcal{C}_{\rho} > 0$, $\mathcal{C}_{\kappa} < 0$ shares many features with 
the one discussed in the main text ($\mathcal{C}_{\rho} < 0$, $\mathcal{C}_{\kappa} > 0$), but it is slightly more involved, hence we have left the discussion for the Additional Material. The high-temperature vev $v_{\kappa}\neq 0$ stemming from $\mathcal{C}_{\kappa} < 0$ in turn leads to a $\rho$-tadpole due to the $\mu$-term in the Zee-Babu scalar potential (Eq.~(1) of the main text). To reabsorb it, a shift in $\rho$ is necessary, generating a non-vanishing vev also for this field. The Universe will then evolve from a high-temperature minimum $(\vev{h}, \vev{\kappa},\vev{\rho}) = (0,v_\kappa,v_{\rho})$, with the hierarchy $v_\kappa \gg v_{\rho}$ granted for $T \gg \mu$ (we moreover observe that this vev hierarchy is generally maintained for $T \lesssim \mu$).
A specific $\{\lambda_i\}$ set corresponding to this setup is given by BM2 in Tab.~\ref{tab1} (which also shows the $\{\lambda_i\}$ values for the benchmark BM1 used in the main text for comparison), with the corresponding numerical results for the allowed parameter space $(m_\kappa, m_\rho)$ shown in Fig.~\ref{fig2:point1}. 
For this benchmark we have $T_c < T_{\rm EW} \simeq  165$ GeV.
We note that the viable parameter space is significantly larger than for BM1, since the possibility of $\Delta T > 0$ now depends on $m_\kappa$, for which the existing LHC constraints are less stringent (recall the discussion in the main text).

\begin{center}
\begin{table}[h!]
\begin{tabular}{||c c c c c c c c c||}
\hline
& $\lambda_{\kappa}$  & $\lambda_{\rho}$ & $\lambda_{h\kappa}$  & $\lambda_{h\rho}$ & $\lambda_{\kappa\rho}$  & $C_h$ & $C_{\kappa}$ & $C_{\rho}$  \\
\hline\hline
BM1 & 0.118 & 5.44 &  4.70 & -0.097 & -0.052 & 0.042 & 0.048 & -0.85  \\
\hline
BM2 & 4.394 & 0.811 & 5.134 & -0.537 & -0.142 & 0.048 & -0.529 & 0.192  \\
\hline
\end{tabular}
\caption{\em \label{tab1}Benchmark scenarios accommodating early Universe hypercharge breaking with $T_{\rm EW} > T_{Y}$, for $\mathcal{C}_{\rho} < 0$, $\mathcal{C}_{\kappa} > 0$ (BM1 - main text) and $\mathcal{C}_{\rho} > 0$, $\mathcal{C}_{\kappa} < 0$ (BM2).}
\end{table}
\end{center}

At this point, we stress that the baryogenesis mechanism for both the setup discussed here and in the main text relies on the dynamical change of lepton masses across the phase transition -- with the simultaneous presence of various lepton mass sources being crucial for the required CP violating effects -- . After the phase transition, the charged vevs vanish and the CP violating sources cease to be active (this bears similarity e.g. with the mechanism discussed recently in~\cite{Huber:2022ndk}), thus rendering the model safe from electric dipole moment (EDM) experimental constraints. Besides, chiral asymmetries across the bubble wall are only generated for leptons, not for quarks (as these only couple to the SM Higgs, and thus there is no complex phase in their mass generated during the transition).\\

Finally, the efficiency of the present baryogenesis mechanism will be sensitive -- as for standard EW baryogenesis setups -- to the expansion velocity of the phase transition bubbles. A precise evaluation of the bubble wall velocity is a highly involved analysis beyond the scope of this Letter, yet we note that the presence of the BSM scalars with sizable couplings to the SM Higgs in the Zee-Babu model is expected to exert significant friction of the bubbles, slowing them down -- and EW baryogenesis setups work best for subsonic wall velocities, see e.g.~\cite{Konstandin:2013caa} -- . Also, we point out that
recent works~\cite{Cline:2020jre,Laurent:2020gpg,Dorsch:2021ubz,Laurent:2022jrs} have shown that supersonic bubble wall velocities can also lead to successful baryogenesis, with a mild suppression of the generated baryon asymmetry compared to that of subsonic velocities.

\section{LHC limits on Zee-Babu singly charged scalars}
\label{AppD}
\begin{figure}[ht]
\begin{centering}
\includegraphics[width=0.495\textwidth]{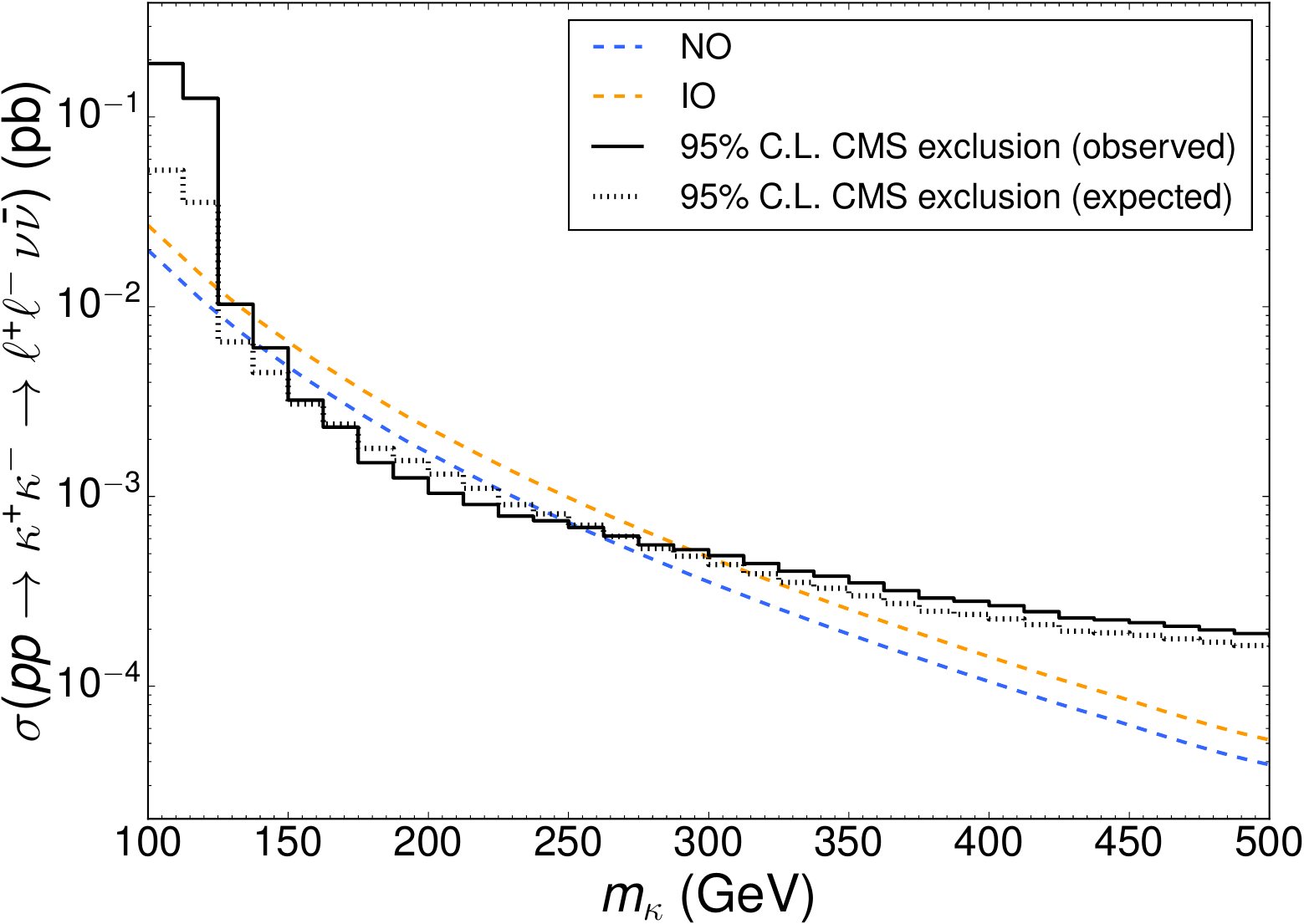} 
\caption{\em $13$ TeV LHC cross section for $p p \to \kappa^{+} \kappa^{-} \to \ell^+ \ell^- \, \nu\bar{\nu}$ ($\ell = e, \mu$) as a function of $m_\kappa$, for NO (dashed blue) and IO (dashed orange). The present 95\% C.L. observed (expected) bounds from CMS~\cite{CMS:2020bfa} are shown as a solid (dotted) black line.}
\label{fig:LHC_limits}
\par\end{centering}
\end{figure}

A direct reinterpretation of the current limits for EW production of right-handed sleptons, each decaying to a SM lepton and a massless neutralino, is possible for the latest CMS analysis with $\sqrt{s} = 13$ TeV c.o.m.~energy and 137 fb$^{-1}$ of integrated luminosity~\cite{CMS:2020bfa}: we first compute the $\kappa^{\pm}$ branching fractions in the ZB model, very approximately (see Sec.~I) given by: BR$(\kappa^{\pm} \to e^{\pm} \nu) = 0.25$, BR$(\kappa^{\pm} \to \mu^{\pm} \nu) =$ BR$(\kappa^{\pm} \to \tau^{\pm} \nu) = 0.375$ for NO and BR$(\kappa^{\pm} \to e^{\pm} \nu) = 0.444$, BR$(\kappa^{\pm} \to \mu^{\pm} \nu) =$ BR$(\kappa^{\pm} \to \tau^{\pm} \nu) = 0.278$ for IO, respectively. We also obtain the $\sqrt{s} = 13$ TeV LHC production cross section for $p p \to \kappa^{+} \kappa^{-}$ (as a function of $m_{\kappa}$) from the equivalent EW pair-production of right-handed sleptons (at NLO and NLL) with \textsc{Resummino}~\cite{Fuks:2013vua}. We can then compare the ZB predicted cross section for $p p \to \kappa^{+} \kappa^{-} \to \ell^+ \ell^- \, \nu\bar{\nu}$ ($\ell = e, \mu$) with the current bounds 95\% C.L. cross section bounds from Fig.~14 of~\cite{CMS:2020bfa} assuming a massless neutralino. The various cross-sections as a function of $m_\kappa$ are shown in Fig.~\ref{fig:LHC_limits}, from which we obtain a 95\% C.L. exclusion for masses in the approximate range $m_{\kappa}\in [150, 250]$ GeV for NO and $m_{\kappa}\in [130, 300]$ GeV for IO.\footnote{Note that the $m_{\kappa}$ bin-size of $12.5$ GeV in Ref.~\cite{CMS:2020bfa} makes the above $m_{\kappa}$ exclusion boundaries only approximate. Furthermore, the public limits~\cite{CMS:2020bfa} apply to the sum of $e^+ e^-$ and $\mu^+ \mu^-$ channels, which in the ZB model have different cross sections. Thus if the experimental sensitivity is not the same for $e^+ e^-$ and $\mu^+ \mu^-$, the actual experimental limit on $m_{\kappa}$ could be somewhat weaker than our derivation.} We have further checked that the current strongest limits from LHC searches for pair-produced right-handed sleptons decaying into $\tau$-leptons~\cite{CMS:2022syk} do not place any constraint on the ZB parameter space. Finally, we also note that a recent theoretical study~\cite{Alcaide:2019kdr} projected an approximate 95\% C.L. bound $m_{\kappa} \gtrsim 500$ GeV for $\sqrt{s} = 13$ TeV LHC with an integrated luminosity of $200$~fb${}^{-1}$, which is however more optimistic than the reinterpretation presented here (yet, ways of improving the sensitivity of these searches have been proposed, see e.g. Ref.~\cite{Fuks:2019iaj}). 

\section{Comparison between Hartree and full thermal potential.}
\label{AppE}
\begin{figure}[h!]
  \centering
    \includegraphics[width=0.495\textwidth]{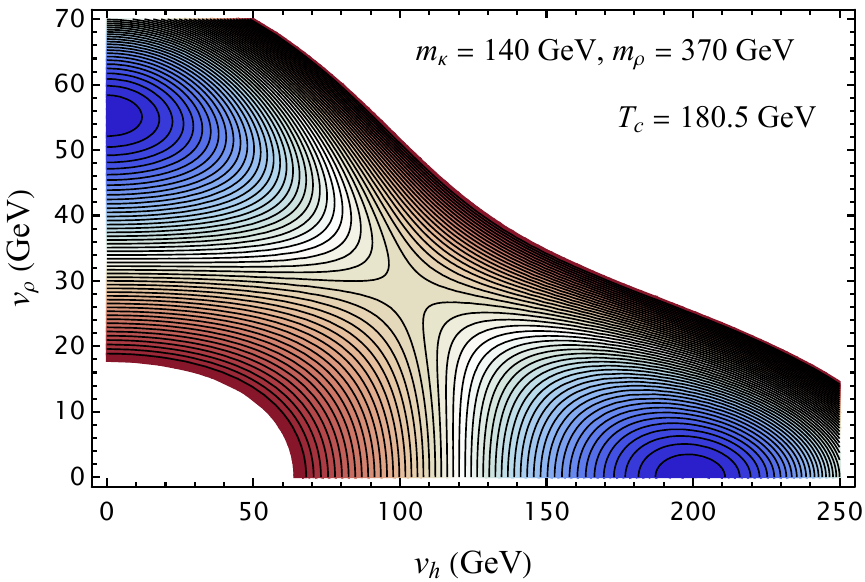}
    \includegraphics[width=0.495\textwidth]{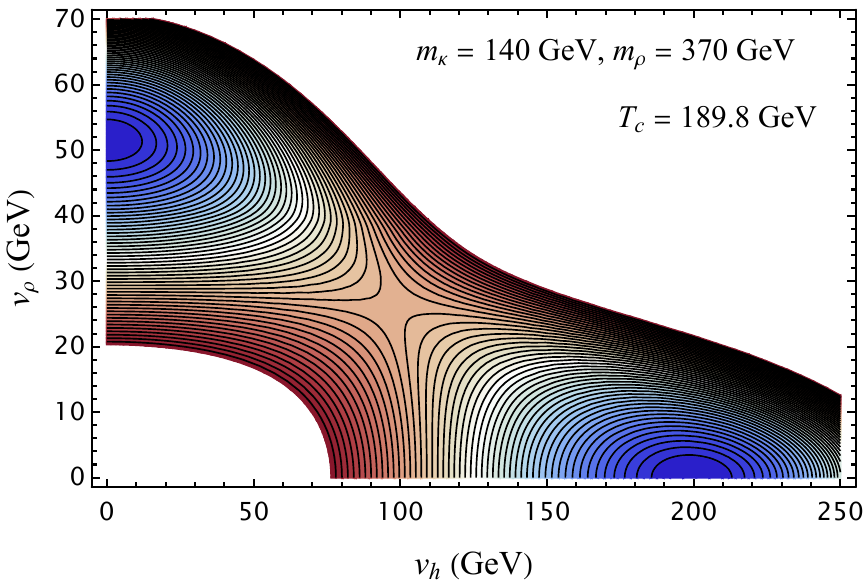}
  \caption{\em Contour plots of the finite-temperature effective potential for BM1 at the critical temperature $T_c$ and fixed scalar masses $m_{\kappa}$, $m_{\rho}$. Top: {\it Hartree} potential. Bottom: full thermal potential approximated by Bessel functions of the second kind to order $l = 5$.}
\label{fig4:fullthermal}
\end{figure}

We give the comparison between the {\it Hartree} and the full 1-loop thermal potential with Daisy resummation. As the exact evaluation of the thermal $J$-functions in the full potential $V_T$ is numerically cumbersome, we employ a well-known expansion in terms of Bessel functions of the second kind \cite{Anderson:1991zb}, i.e.
\begin{equation}
    J_{\pm} \left( x \right) = \lim_{l\to +\infty} \pm \sum_{n=1}^{\l} \frac{(\mp 1 )^n x}{n^2} K_2(\sqrt{x}n) \, .
\end{equation}
This expansion exhibits an excellent convergence behaviour towards the exact functions over a wide range of $x$ for as low as $l = 5$~\cite{Bernon:2017jgv}, thus covering a vast temperature range. For the exact evaluation of the Daisy contribution $V_{\rm D}$ (see Sect.~III in the main text) we employ the full thermal masses for the $Z$ and $\gamma$ given in Eq.~\eqref{fullthermalmasses} instead of the high-temperature approximation of Eq.~\eqref{fullthermalmasse3} used throughout this work. In Fig.~\ref{fig4:fullthermal} we proceed to benchmark the goodness of the truncated {\it Hartree} approximation (top) against the full potential (bottom) on the basis of BM1. On these contour plots at fixed scalar masses $m_{\kappa}$ and $m_{\rho}$ the {\it Hartree} approximation displays the general shape and phase structure of the full potential, with high- and low-temperature minima well separated by a barrier suggesting a first-order phase transition. We also note that the vevs of the high- and low-temperature minima as well as the critical temperature $T_c$ are reproduced within reasonable bounds. This establishes the ability of the {\it Hartree} approximation to capture details of the thermal history crucial to our results such as the type and strength of the phase transition.

\providecommand{\href}[2]{#2}
\begingroup\raggedright\endgroup

\end{document}